\documentclass{aa}
\usepackage{psfig}

\begin{document}

   \thesaurus{06         % A&A Section 6: Form. struct. and evolut. of stars
              (19.53.1;  % Stars: oscillations of,
               19.63.1)} % Stars: structure of.
   \title{The distance modulus of the Large Magellanic Cloud based on 
          double-mode RR Lyrae stars}

   \subtitle{}

   \author{G. Kov\'acs
          \inst{}
%          \and
%          C. Ptolemy\inst{2}\fnmsep\thanks{Just to show the usage
%          of the elements in the author field}
          }

   \offprints{G. Kov\'acs}

   \institute{Konkoly Observatory, P.O. Box 67, H-1525, Budapest, Hungary \\
              email: kovacs@konkoly.hu
%         \and
%             University of Alexandria, Department of Geography\\
%             email: c.ptolemy@hipparch.uheaven.space
%             \thanks{The university of heaven temporarily does not
%                     accept e-mails}
             }

   \date{Received ...; accepted ...}

   \maketitle

   \begin{abstract}

We use double-mode RR Lyrae (RRd) stars from the {\sc macho} variable 
star database of the Large Magellanic Cloud (LMC) to estimate its 
distance, by utilizing photometric data, linear pulsation and stellar 
atmosphere models. If we set $E_{B-V}=0.11$ and [M/H]$=-1.5$ for LMC, 
we get $M-m=18.52$~mag, which is in agreement with the value obtained 
from Galactic cluster RRd stars. Although the formal statistical error 
(standard deviation) of $M-m$ derived from the {\sc macho} set is only 
$0.024$~mag, the systematic errors are more substantial. These errors 
are mostly due to the zero point ambiguity of the temperature scale and 
to the error of the photometric calibration. However, the fact that the 
distance moduli obtained in this and in our former studies of cluster 
RRd and Small Magellanic Cloud beat Cepheids agree so well, implies that 
the only serious source of error is the zero point of the temperature 
scale, which should not have larger than $\pm 0.10$~mag effect on the 
distance modulus.  
   

   \end{abstract}

%
%________________________________________________________________

%%%%%%%%%%%%%%%%%%%%%%%
%%     SECTION 1     %%
%%%%%%%%%%%%%%%%%%%%%%%

\section{Introduction}

The determination of the distance moduli ($DM$) of the Magellanic Clouds 
plays an important role in establishing the cosmic distance scale. 
Mostly due to the analyses of the {\sc hipparcos} data, there is a 
renewed effort to pin down this crucial distance within an accuracy 
of better than $\approx 0.05$~mag. However, the question is far from 
being settled, and the LMC distance moduli obtained with 
various methods may differ by several tenth of magnitudes, 
stretching from the `short' ($DM\approx 18.2$, e.g., Stanek et al. 2000) 
to the `long' ($DM\approx18.6$, e.g., Feast 1999) distance scales. 

In the first paper of this series (Kov\'acs \& Walker 1999, hereafter 
KW99, see also Kov\'acs 2000b) we used RRd stars in Galactic clusters 
to show that they yield $0.2-0.3$~mag larger distances than the ones 
inferred from the Baade-Wesselink absolute magnitudes of RR~Lyrae 
stars. This result leads to $DM\approx 18.5$~mag for the LMC. When the 
same method is applied to the Small Magellanic Cloud (SMC) beat Cepheids 
with 0.51~mag relative distance between the SMC and LMC, we get almost 
the same $DM$ for the LMC (Kov\'acs 2000a). In the present Letter we 
directly extend the analysis to the {\sc macho} RRd stars (Alcock et al. 
1997, 2000) and show that these give larger distance as well, in fine 
agreement with the above mentioned independent data sets.

%%%%%%%%%%%%%%%%%%%%%%%
%%     SECTION 2     %%
%%%%%%%%%%%%%%%%%%%%%%%

\section{Method, data, parameters}

We use the same method as in KW99. This implies the utilization of: 
(a) multicolor data and reddening values on the LMC RRd variables;  
(b) linear pulsation models; 
(c) color~$\rightarrow$~$T_{\rm eff}$ transformations given by 
the static stellar atmosphere models and by the InfraRed Flux Method 
({\sc irfm}). When considering cluster RRd variables in the context 
of the LMC distance modulus, we need the same types of information, 
in addition to the knowledge of the relative cluster distances. 

Just briefly mentioning the essence of the method, and referring to 
KW99 for further details, we note that the pulsation equations yield 
a relation for the luminosity as a function of the periods, temperature 
and metal abundance 
%
%>>>>>>>>>>>>>>>>>>>>>>>>>>>>>>>
%    EQ. (1)
%>>>>>>>>>>>>>>>>>>>>>>>>>>>>>>>
%
\begin{eqnarray}
\log L = f(P_0, P_1, \log T_{\rm eff}, {\rm [M/H]}) \hskip 2mm .
\end{eqnarray}
\noindent
There is also a dependence on other details of the chemical 
composition, but it has been proved to be rather weak at 
the relatively low metallicities of the RRd stars in our sample. 
From the observed dereddened colors we can get estimates on $T_{\rm eff}$. 
The metal content [M/H] is preferably obtained from direct spectroscopic 
observations, but we may also assume that RRd and fundamental mode (RRab) variables have the same [M/H] in a given cluster and apply the method 
of Jur\-csik \& Kov\'acs (1996) to determine the 
metallicity of the RRab stars. Once the luminosity is computed, using 
the bolometric correction formula of KW99 simply leads to the distance 
modulus through the comparison with the observed magnitudes.   

In the following we give a more detailed description of the parameters 
entering in the calculation of the LMC distance modulus.  

Together with the earlier discoveries of Alcock et al. (1997), the 
recent analysis of more than 1300 short-period RR~Lyrae stars led 
to a substantial extension of the known RRd stars in the LMC 
(Alcock et al. 2000). In the present analysis we include 181 RRd variables, 
which constitute {\it all} presently known RRd stars in the LMC. 
The {\sc macho} instrumental magnitudes have been transformed to the 
standard Johnson $V$ and Kron-Cousins $R_c$ colors according to the 
recipe of Alcock et al. (1999). The periods and the average magnitudes 
are listed in Table~1.\footnote{Table~1 is available only in electronic 
form at CDS (ftp 130.79.128.5)} 

For an independent computation of the LMC distance modulus we use globular 
cluster RRd data compiled by KW99. Relative distance moduli required 
by this method are checked by the photometric data of Udalski (1998) 
and Clementini et al. (2000b, hereafter C00b).   

The $T_{\rm eff}=f({\rm color}, \log g, {\rm [M/H]})$ relations 
are derived from the stellar atmosphere models of Castelli et al. 
(1997, hereafter C97) with the zero point adjusted to the {\sc irfm} 
results of Blackwell \& Lynas-Gray (1994, hereafter BLG94). The 
required color, $\log g$ and [M/H] data are obtained from Clementini 
et al. (1995). It is important to remark that this approach assumes 
a uniform shift in $\log T_{\rm eff}$ between the {\sc irfm} and 
theoretical scales, and that this shift is applicable throughout the 
relevant parameter regime (i.e., from dwarfs to giants). Additional 
difficulty might occur because of the inaccuracy of the [Fe/H] and 
$\log g$ values given by the {\sc irfm} sources, or the neglect of 
the $\log g$ dependence in some of those works (e.g., Alonso et al. 
1996; 1999, hereafter A96 and A99, respectively). Furthermore, 
different colors may yield different shifts, producing inconsistency 
among the derived temperatures at a level of $0.004$ in $\log T_{\rm eff}$. 
By considering various colors and overlapping samples in the {\sc irfm} 
publications, our current estimates for the zero point differences 
(in the sense of $\log T_{\rm eff}$~(source) minus $\log T_{\rm eff}$~(C97)) 
are as follows: $-0.004$~(BLG94), $-0.007$~(A96); 
$-0.008$~(Blackwell \& Lynas-Gray 1998); $-0.010$~(A99). Here we use 
the scale of BLG94, because it is close to the one used in our previous 
studies and in the Baade-Wesselink analyses (see KW99). In Sect. 3 we 
will discuss the effect of the $T_{\rm eff}$ zero point on the distance 
determination. The final formulae, adjusted to the BLG94 scale are the 
following 
%   
%>>>>>>>>>>>>>>>>>>>>>>>>>>>>>>>
%    EQS. (2), (3)
%>>>>>>>>>>>>>>>>>>>>>>>>>>>>>>>
%
\begin{eqnarray}
\log T_{\rm eff} & = & 3.8804 - 0.3213(B-V) + 0.0176\log g \nonumber \\ 
             & + & 0.0066{\rm [M/H]} \hskip 2mm , \\
\log T_{\rm eff} & = & 3.8928 - 0.4910(V-R_c) + 0.0116\log g \nonumber \\
             & + & 0.0012{\rm [M/H]} \hskip 2mm . 
\end{eqnarray}
\noindent
As far as the metallicities are concerned, for the three Galactic 
clusters (M15, M68 and IC4499) we use the data given in KW99. For 
the LMC field there are very recent direct $\Delta S$ measurements 
by Clementini et al. (2000a). The average [Fe/H] value of their 7 RRd 
variables is $-1.6$ (or $-1.5$ if the less accurate low-metallicity 
star is left out). Here we accept ${\rm [Fe/H]}=-1.5$ for the LMC 
RRd variables. In principle, one should use variable dependent 
metallicity as it follows from the analysis of the period ratio 
diagram (Popielski et al. 2000). However, tests have shown that in 
the present context and data quality, this dependence has a rather 
negligible effect on the final conclusion. Therefore, we use the same 
metal abundance for all variables in the LMC. 

In order to compare the direct $DM$ obtained from the LMC field RRd 
variables with the ones computed from cluster RRd stars, we have 
to know the relative distances and reddenings. In calculating these 
quantities we follow essentially the method of Kov\'acs \& Jurcsik 
(1997) which employs the periods and the Fourier decompositions of 
the light curves in computing the physical parameters of RRab stars. 
In the present implementation of the method we use only the period 
in deriving the relations for the reddening-free quantities 
(e.g., $W=V-3.1(B-V)$, $(B-V)_0$). This approximation is necessary, 
because of the limited information available on the LMC {\sc ogle} 
data set. Nevertheless, the results are sufficiently accurate for 
the present purpose (see also Kov\'acs 
\& Walker 2000). 

It is also possible to make a direct comparison between the colors 
and distance moduli of the RRd variables in LMC and IC4499. Since 
they both have very similar metallicities and periods, we may assume 
that their average intrinsic colors and luminosities are also very similar.
The average quantities for the 181 {\sc macho} LMC RRd stars are the 
following: $\langle V \rangle = 19.30$, $\langle V-R \rangle = 0.246$, 
$\langle P_0 \rangle = 0.49$. Similarly, from C00b for the 10 RRd stars 
we get: 
$\langle V \rangle = 19.29$, $\langle B-V \rangle = 0.37$, 
$\langle P_0 \rangle = 0.48$.
The same quantities for the 15 RRd stars in IC4499 are the following: 
$\langle V \rangle = 17.66$, $\langle B-V \rangle = 0.50$,  
$\langle V-R \rangle = 0.31$, $\langle P_0 \rangle = 0.48$. Simple 
subtraction of the corresponding quantities leads to the results 
summarized in Table 2. 

The close match between the relative distance moduli and reddenings 
calculated from the different data sets is very encouraging as far 
as the consistency among the photometric zero points are concerned. 
We note in passing that relative distance moduli calculated by the 
method of Kov\'acs \& Jurcsik (1997) from the RRab variables of 
four LMC globular clusters also yield 2.04~mag for the average of the 
relative distances between those LMC clusters and IC4499.  
%
%>>>>>>>>>>>>>>>>>>> Table 2.
%
\setcounter{table}{1}
\begin{table}[h]
\caption[ ]{Relative distance moduli (LMC minus IC4499) and reddenings 
of LMC field RR Lyrae stars}
\begin{flushleft}
\begin{tabular}{lrrr}
\hline
Source   & $N$ & $\Delta DM$  & $E_{B-V}$    \\
\hline
Field ({\sc ogle} RRab): & 100 & 2.04 & 0.13 \\
RRd({\sc macho}):        & 181 & 1.99 & 0.11 \\
RRd(C00b):               &  10 & 2.03 & 0.09 \\
\hline
\end{tabular}
\end{flushleft}
\end{table}
%
%

%%%%%%%%%%%%%%%%%%%%%%%
%%     SECTION 3     %%
%%%%%%%%%%%%%%%%%%%%%%%

\section{Results}

In the following we derive the distance of the LMC from two independent 
data sets containing RRd stars: (1) Galactic globular clusters; 
(2) {\sc macho} LMC field variables. We use composition (d) of KW99 
($X=0.76$, with solar-type distribution of heavy elements), and 
{\sc opal'96} opacities (Iglesias \& Rogers 1996). 

Relative distance moduli between the three Galactic globular clusters 
and their reddenings are from KW99. These parameters have been calculated 
by the method of Kov\'acs \& Jurcsik (1997). We fix the relative distance 
modulus between LMC and IC4499 at $2.03$~mag and the LMC reddening at 
$E_{B-V}=0.11$. Column 4 of Table 3 shows the derived distance moduli. 
The agreement is very good between the different estimates. This shows 
that the relative values of all parameters have been chosen properly. 
Our more serious concern is the effect of the various zero point shifts. 
In the subsequent columns of Table 3 we show the changes of $DM$ due to 
these shifts and also to those ambiguities which occur in the pulsation 
models. Except for the case of changing [M/H], negative shifts in the 
parameters cause the same amount of negative offsets in $DM$. If [M/H] 
is decreased, the corresponding changes are up to 0.04~mag smaller in 
$DM$ than if [M/H] is increased by the same amount. The adopted values 
for the changes of the tested quantities are merely to demonstrate their 
effects without implying that these changes can, in fact, be expected. 
In particular, it is rather unlikely that for Galactic globular clusters 
the reddenings were off by $0.03$~mag or more. In any case, zero point 
effects in $E_{B-V}$ are not serious in our method, because they are 
compensated in the derived $DM$ through the nearly parallel effects 
in the calculated luminosity and in the magnitude corrections for 
reddening (see also KW99). 

We see that almost all zero point errors cause uniform shifts in the 
distance moduli, which are independent of the stellar population. The 
only exception is the zero point of the metallicity scale. Variables 
with higher [M/H] are more affected by this zero point ambiguity. 
Therefore, metallicity change in itself cannot decrease the 
distance modulus too much, because this would result in a substantial 
increase of the differences between the computed distance moduli. 

Ambiguity in the zero point $T_0$ of the color~$\rightarrow T_{\rm eff}$ 
transformation is the main source of error in the present method. 
However, we note that even if we accepted the temperature scale of A99, 
which is the lowest among the available {\sc irfm} scales, the 
derived distances would be still around $\approx 18.4$ mag. We are not 
aware of any studies strongly indicating the possibility of more 
dramatic lowering of the temperature scale. We note that the method of 
color averaging might have also some effect on the derived effective 
temperature. Here we use magnitude averaging, which has a tendency 
of yielding redder equilibrium colors, i.e., lower temperatures than 
the static ones (Bono et al. 1995). 

Effect of period ratio change due to nonlinearity is rather small. 
Past and very recent works on convective RRd models show that period 
ratio changes are within $0.002$. In the case of RRd models, there is 
a trend of the nonlinear convective period ratios being smaller than the 
linear radiative ones (Koll\'ath \& Buchler 2000). This would result in 
a slight {\it increase} in the derived $DM$, because our results are 
based on radiative linear models, but on the other hand, nonlinear period 
increase might compensate for this effect.  

In the last column we show the small change due to varying hydrogen 
content from $X=0.76$ to $0.70$ (mixtures (c) and (d) in KW99). 
% 
%
%>>>>>>>>>>>>>>>>>>> Table 3.
%
\begin{table*}[t]
\caption[ ]{Derived distance moduli for the LMC. Columns 5--8 show the 
changes in the distance modulus, if $E_{B-V}$, [M/H], 
$\log T_0$, $P_1/P_0$~(observed) are changed by $+0.03$, $+0.2$, $+0.005$ 
and $+0.001$, respectively. Column 9 shows the effect of changing 
$X$ from $0.76$ to $0.70$  
}
\begin{flushleft}
\begin{tabular}{lcccccccc}
\hline
Cluster   & [M/H]  & $E_{B-V}$  & $DM_{LMC}$ 
          & $\Delta DM_{E_{B-V}}$ 
          & $\Delta DM_{\rm [M/H]}$ 
          & $\Delta DM_{\log T_0}$ 
          & $\Delta DM_{P_1/P_0}$ 
          & $\Delta DM_{Comp.}$  \\
\hline
M15     & $-2.3$ & 0.07 & 18.52 & 0.04 & 0.05 & 0.07 & 0.05 & 0.01 \\
M68     & $-2.0$ & 0.03 & 18.47 & 0.05 & 0.08 & 0.07 & 0.06 & 0.02 \\
IC4499  & $-1.5$ & 0.22 & 18.50 & 0.06 & 0.15 & 0.08 & 0.05 & 0.04 \\
LMC     & $-1.5$ & 0.11 & 18.52 & 0.04 & 0.14 & 0.08 & 0.06 & 0.04 \\
\hline
\end{tabular}
\end{flushleft}
\end{table*}
%
%
%>>>>>>>>>>>>>>>>>>>>
%
%   FIGURE 1.
%
%>>>>>>>>>>>>>>>>>>>>
%
\begin{figure}[t]
\vskip 1mm
\centerline{\psfig{figure=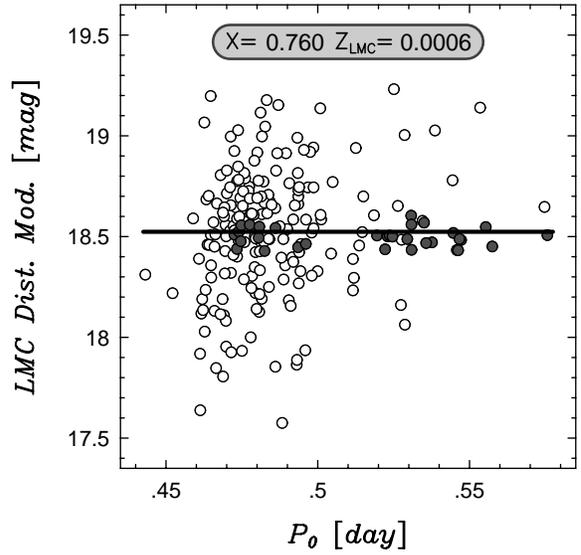,height=75mm,width=70mm}}
\vskip -0mm
\caption{
Individual distance moduli from the LMC (open circles) and from 
Galactic RRd stars (gray circles). The horizontal line shows the 
average distance modulus derived from the LMC variables 
}
%%\vskip 0mm
\end{figure}
Finally, to exhibit the distribution of the distance moduli derived 
for the individual stars, Fig. 1 shows these values as a function of 
the period. The huge scatter of the LMC stars is partially counterbalanced 
by their large number, which brings the statistical error down to 
$0.024$~mag for the distance modulus derived from these variables. 
This can be compared with the value of $0.015$~mag, valid for the 
Galactic RRd stars. The largest contribution to the $\sigma(DM)=0.32$~mag 
scatter of the LMC stars comes from the photometric errors. By using the 
error estimates of $\sigma_V=0.02$, $\sigma_{(V-R)}=0.03$ (see Alcock et al. 1999) we get 0.21~mag for the standard deviation of the individual 
distance moduli. We note that the result is most sensitive to 
$\sigma_{(V-R)}$. For example, with $\sigma_{(V-R)}=0.04$ we get   
$\sigma(DM)=0.26$~mag. The rest of the scatter comes from the metallicity 
dispersion, inhomogeneous reddening and possibly from the spatial extent  
of the LMC. 

%%%%%%%%%%%%%%%%%%%%%%%
%%     SECTION 4     %%
%%%%%%%%%%%%%%%%%%%%%%%

\section{Conclusions}

The recent revision of the {\sc macho} RRd magnitudes (Alcock et al. 
1999) enabled us to calculate the distance of the LMC directly from the 
observed magnitudes, colors and periods. This shows a very good agreement 
with the independent estimate based on Galactic globular cluster RRd stars 
and SMC beat Cepheids (Kov\'acs 2000a). They all give $M-m\approx18.5$~mag, 
which lends further support to the `long' distance scale. The largest 
source of systematic error is the potential ambiguity in the zero point 
of the temperature scale. Even this is not likely to cause the large 
effect, which would be necessary to lower the distance modulus by 
$\approx 0.3$~mag, the value required by some current estimates based on 
other methods (e.g., Stanek et al. 2000). This amount of change would 
demand about $300$K lower temperature scale than the one used in this 
work. Since our adopted scale is only 100~K hotter than the presently 
published lowest infrared flux scale of Alonso et al. (1999), we think 
that it is highly unlikely that the required lowering of the distance 
modulus is possible by temperature change alone. Consequently, we would 
need some `conspiracy' with metallicity, reddening or other parameters, 
so as to get the necessary decrease in the distance modulus. This 
situation is also not very probable, because, although the strongest 
effect comes from the metallicity, this is primarily differential i.e., 
the distance moduli derived from the low metallicity clusters would not 
be seriously affected. This would lead to inconsistency among the 
distance moduli, because the relative distances, being empirically 
determined, are independent from the adopted metallicities. It is 
also worthwhile to mention that the metallicity values used in this and 
in our previous studies are in very good agreement with the ones published 
by independent spectroscopic studies and that these metallicities minimize 
the differences between the LMC distance moduli derived from various 
double-mode stars. 

We conclude that the application of standard temperature and metallicity 
scales leads to the `long' RR~Lyrae distance scale, which, except for 
a drastic lowering of the temperature scale, is not possible to match 
with the `short' distance scale.    

%%%%%%%%%%%%%%%%%%%%%%%%%%%%%%
%%     ACKNOWLEDGEMENTS     %%
%%%%%%%%%%%%%%%%%%%%%%%%%%%%%%

\begin{acknowledgements}
This work was started during the author's stay at the Lawrence Livermore 
Laboratory in 1998. He is very grateful for the hospitality of Kem Cook 
and Charles Alcock. Special thanks are due to Kem Cook for his help in  
transforming the data to the standard system. Critical comments by David 
Alves on the draft version of this Letter are thanked. Correspondence on 
the LMC photometry with Andrzej Udalski and discussions on the nonlinear 
period changes with Zolt\'an Koll\'ath are very much appreciated. A part 
of this work was done during the author's stay at the Copernicus 
Astronomical Center in Warsaw. Hospitality and useful discussions with 
Woj\-tek Dziembowski, Pawel Moskalik and Blazej Popielski are thanked. 
Useful discussions with B\'ela Szeidl and G\'asp\'ar Bakos were also 
very instrumental in the near completion phase of this work. The supports 
of the following grants are acknowledged: {\sc otka t$-$024022, t$-$026031} 
and {\sc t$-$030954}.
\end{acknowledgements}

\end{document}